\documentclass[a4paper,11pt]{article}
\pdfoutput=1 

\usepackage{jcappub}

\title{\boldmath Do sewn singularities falsify the Palatini cosmology?}

\author[a,b]{Marek Szyd{\l}owski}
\author[a]{Aleksander Stachowski}
\author[c]{Andrzej Borowiec}
\author[c,d]{Aneta Wojnar}

\affiliation[a]{Astronomical Observatory, Jagiellonian University, Orla 171, 30-244 Krakow, Poland}
\affiliation[b]{Mark Kac Complex Systems Research Centre, Jagiellonian University, ul. {\L}ojasiewicza 11, 30-348 Krak{\'o}w, Poland}
\affiliation[c]{Institute for Theoretical Physics, University of Wroc{\l}aw, pl. M. Borna 9, 50-204, Wroc{\l}aw, Poland}
\affiliation[d]{Dipartimento di Fisica, Universita’ di Napoli Federico II,
Complesso Universitario di Monte S. Angelo, Via Cinthia, 9,
I-⁠80126 Naples, Italy
}

\emailAdd{marek.szydlowski@uj.edu.pl}
\emailAdd{aleksander.stachowski@uj.edu.pl}
\emailAdd{andrzej.borowiec@ift.uni.wroc.pl}
\emailAdd{aneta.wojnar@ift.uni.wroc.pl}

\abstract{
We investigate further (cf. \href{http://arxiv.org/pdf/1512.01199.pdf}{JCAP01 (2016) 040}) Starobinsky cosmological model $R+\gamma R^2$ in the Palatini formalism with Chaplygin gas and baryonic matter as a source. For this aim we use dynamical system theory. The dynamics is reduced to the 2D sewn dynamical system of a Newtonian type (a piecewise-smooth dynamical system). We classify all evolutional paths in the model as well as trajectories in the phase space. We demonstrate the presence of a degenerate freeze singularity (glued freeze type singularities) for the positive $\gamma$. In this case it is a generic feature of early evolution of the universe. We point out that a degenerate type III of singularity can be considered as an endogenous model of inflation between the matter dominating epoch and the dark energy phase. We also investigate cosmological models with negative $\gamma$. It is demonstrated that $\gamma$ equal zero is a bifurcation parameter and dynamics qualitatively changes in comparison to positive $\gamma$. Instead of the big bang the sudden bounce singularity of a finite scale factor appears and there is a generic class of bouncing solutions sewn along the line $a=a_{\text{sing}}$. And we argue that the presence of sudden singularities in an evolutional scenario of the Universe falsifies the negative $\gamma$ in the Palatini cosmology. Only very small values of $\Omega_{\gamma}$ parameter are admissible if we requires that agreements physics with the $\Lambda$CDM model. From the statistical analysis of astronomical observations, we deduce that the case of negative values of $\Omega_\gamma$ can be rejected even if it may fit better to the data.}

\begin{document}
\maketitle

\section{Introduction: Cosmology with Chaplygin gas in Palatini formalism}

Today's modern cosmology suffers problems which the standard theory, that is the $\Lambda$CDM model derived from Einstein's general relativity, is not able to explain. There are such issues like dark matter and dark energy, origin and source of inflation or large scale structure which are widely investigated from many different points of view. Satisfactory explanations have not been set so far which makes searching and examination of new models still desirable. Due to that fact we have proposed \cite{Borowiec:2015qrp} and examined a model which modifies the standard one in two different ways. The first modification refers to the gravitational Lagrangian which we enriched with the so-called Starobinsky term $\gamma\hat{R}^2$ keeping in mind that the parameter $\gamma$ is small and being about to estimate by observational data and analytical analysis. We have used the Palatini formalism \cite{Palatini:1919di,DeFelice:2010aj,Capozziello:2011et}, which treats a metric and a connection as independent dynamical variables. The connection is used to construct Riemann and Ricci tensors, while contraction with the metric provides (generalized) Ricci scalar (see e.g. \cite{DeFelice:2010aj,Sotiriou:2008rp}).

Palatini approach to the description of gravitational field was originally introduced by Einstein himself \cite[p.~415]{Capozziello:2011beg} but historical misunderstanding decides its name in this context \cite[p.~191, 485]{Capozziello:2011beg}. The main idea of this formalism is to treated the connection $\Gamma$ appeared in the definition of the Ricci tensor as a variable independent of the spacetime metric $g$. Therefore there is no special reasons to apply the Palatini variational principle in GR if we have metric formalism. The situation changes if we considered Extended Theory of Gravity (ETG) because in these cases both metric and Palatini variational principle satisfy different field equations which in order can give rise to different physics \cite[p.~486, 769]{Capozziello:2011beg}. Application of the Palatini approach in the context of cosmological investigations has been the subject of many papers \cite[p. 211, 721, 722, 843, 1129]{Capozziello:2011beg}. The Newtonian potential can be obtained if we considered the weak-field limit of ETG and its relations with a conformal factor \cite[p.~794]{Capozziello:2011beg}. In particular, it has been shown \cite{Ferraris:1992dx,Borowiec:1996kg} that vacuum solutions of Palatini gravity differ from GR by the presence of cosmological constant. Due to this fact the values of cosmological constant which are admitted by solar system tests are many order of magnitude bigger than the values
obtained from cosmological estimations \cite{Allemandi:2006bm}.

Palatini approach is very important in cosmology as the equations of motion are the second order differential equations in comparison to the standard metric approach in which the field equations give rise to the fourth order ones (see e.g. \cite{Borowiec:2015qrp} and references therein).
The second modification was taken to the matter part of the modified Einstein equation. Instead of considering
perfect fluid with barotropic equation of state $p=\omega\rho$ we have studied the so-called generalized
Chaplygin gas which has also, similarly to cosmological constant, a negative pressure. It has gained a lot of attention in cosmology recently
\cite{Bento:2002ps,Kamenshchik:2001cp,Lu:2009zzm,Bilic:2001cg,Popov:2009bt,Naji:2014kaa,Kremer:2004bf,Gorini:2002kf,Avelino:2014nva,Kahya:2015dpa,Fabris:2010vd}
as it combines dark energy (cosmological constant) and dark matter into one component. Moreover, this is the only fluid known up to now which has a supersymmetric
generalization \cite{Hoppe:1993gz,Jackiw:2000cc}. It seems to be a very important and interesting object to study.
We would like to mention that in \cite{Borowiec:2015qrp} we have obtained a very good agreement of our model with observational data. We were also able to find an upper bound on the value of the parameter $\gamma$ in order to locate a singularity, which has appeared in the model, before the recombination epoch. It turn out that the singularity is of the type III and provides an intermediate inflation phase to the evolution of our Universe. That is, the model provides four cosmic evolution phases while the Big Bang singularity is preserved: the decelerating phase dominated by matter, an intermediate inflation phase corresponding to the type III singularity, a phase of matter domination (decelerating phase) and finally, the phase of acceleration of the current universe.
Now on, for the reader convenience, let us shortly remind some properties of the approach that we are using.

The general action of the Palatini $f(R)$-gravity is written in the standard way
\begin{equation}
S=\frac{1}{2}\int d^4 x \sqrt{-g}f(\hat{R})+S_{\text{m}},
\end{equation}
where $f(\hat{R})$ is function of the Ricci scalar $\hat{R}=g^{\mu\nu}\hat{R}_{\mu\nu}(\hat{\Gamma})$. One should notice that the Palatini scalar $\hat{R}$
is constructed with both objects, that is, the metric and connection. The
action $S_{\text{m}}$ is a matter action which is independent of the connection but includes other scalar fields and depends on $g_{\mu\nu}$. Because the Lagrangian for matter does not depend on the connection \cite{Borowiec:2015qrp} in the our model with barotropic matter satisfying equation of state $p=p(\rho)$ can be postulated like in GR in the following form \cite{Minazzoli:2012md}
\begin{equation}
\mathcal{L}_{\text{m}}=-\rho\left(1+\int\frac{p(\rho)}{\rho^2}d\rho\right).
\end{equation}
 Variation of the total action with respect to the metric gives rise to the following field equations
\begin{equation}
f'(\hat{R})\hat{R}_{\mu\nu}-\frac{1}{2}f(\hat{R})g_{\mu\nu}=T_{\mu\nu},\label{structural}
\end{equation}
 where prime denotes differentiation with respect to $\hat{R}$ while the energy-momentum tensor $T_{\mu\nu}$ is obtained by the variation of
 the matter action with respect to  $g_{\mu\nu}$. We have also used the geometric units $8\pi G=c=1$.
 After taking $g$ -trace of (\ref{structural}) one obtains the structural equation in the form
\begin{equation}
f'(\hat{R})\hat{R}-2 f(\hat{R})=T,\label{structural2}
\end{equation}
where $T$ is trace of the energy-momentum tensor. We will discuss that equations a bit later.

The variation with respect to the independent connection $\hat{\Gamma}$ provides
\begin{equation}
 \hat{\nabla}_\lambda(\sqrt{-g}f'(\hat{R})g^{\mu\nu})=0
\end{equation}
which is the indication that the connection is the Levi-Civita connection of the metric $h_{\mu\nu}=f'(\hat{R})g_{\mu\nu}$. That is, the metric $h_{\mu\nu}$
is a conformally related metric to the physical metric $g_{\mu\nu}$. It gives rise to the conclusion that the conformal
factor $f'(\hat{R})$, later labeled by $b$, must be a non-negative function. One should also notice that in the scalar-tensor representation of the
$f(\hat{R})$ gravity in the Palatini approach the scalar field $\phi$ is represented by $f'(\hat{R})$. Moreover, converting the action to the
Einstein frame one gets the negative coupling to the matter part for the case of $f'(\hat{R})<0$.

Equation~(\ref{structural2}) for different choices of function $f(\hat{R})$ leads to the algebraic equation depending on $\hat{R}$. Such an algebraic
equation may happen to be difficult to solve. Postulating the dependence $\hat{R}(T)$, the structural equation becomes a linear differential equation. If
we put $\hat{R}(T)=-T$ in an analogous way to general relativity then we obtain differential equation in the form
\begin{equation}
 T\frac{df(T)}{dT} - 2f(T) = T.
\end{equation}
One gets immediately a simple solution
\begin{equation}
 f(T)=\text{const}\; T^2 - T \quad \text{or} \quad f(\hat{R})=\hat{R}+\text{const}\;\hat{R}^2
\end{equation}
and therefore this class of functions $f(\hat{R})$ defines a full range of choices basing on an analogy to general relativity.

The energy-momentum tensor $T_{\mu\nu}$ satisfies the metric covariant conservation law $\nabla^{\mu} T_{\mu\nu}=0$ since we are considering
the Palatini $f(\hat{R})$ gravity as a metric theory~\cite{Sotiriou:2008rp}. Hence, the continuity equation is given
\begin{equation}
\dot{\rho}+3H(\rho+p)=0,\label{continuity}
\end{equation}
where $\dot{ }\equiv \frac{d}{dt}$ denotes the differentiation with respect the cosmological time, $\rho$ and $p$ are energy density and
pressure, respectively, of the energy momentum for the perfect fluid. The variable $H=\frac{d(\log a)}{dt}$ denotes as usually the Hubble parameter.
For completeness, the form of the equation
of state $p=p(\rho)$ should be postulated in order to obtain the scale factor $a(t)$ dependence for the pressure and energy density. Due to that fact, we
would like to consider a very special case which is perfect fluid for the generalized Chaplygin gas as a source of gravity~\cite{Kamenshchik:2001cp}.
The equations of state is taken as
\begin{equation}
p=-\frac{A}{\rho^{\alpha}},\label{p}
\end{equation}
where $A$ is positive constant and $0\leq\alpha\leq 1$.
Note that negative pressure of generalized Chaplygin gas, for $A>0$, suggests naively that as $\rho$ goes to zero, $p$ diverges to minus infinity. But this reasoning is not true. The pressure as well as energy density satisfy the continuity condition which gives rise to the relation
\begin{equation}
\rho=\rho(a)=\left(A+\frac{B}{a^{3(1+\alpha)}}\right)^\frac{1}{1+\alpha}. \label{rho}
\end{equation}
We parametrize $\rho(a)$ dependence through the physical parameters $\rho_{\text{ch},0}$ and assume $0\leq A_{\text{s}}\leq 1$ (and $B > 0$) \footnote{In some models of fluid inflation  negative values are also allowed \cite{Kahya:2015pza}.} following Bento et al. parametrization, where $A_{\text{s}}=\frac{A}{\rho_{\text{ch,0}}^{1+\alpha}}$ \cite{Bento:2002ps}. In this parametrization the square of speed of sound $\alpha \rho_{\text{ch},0}^{1+\alpha} A_s \rho_{\text{ch}}^{-(\alpha +1)}$, today $c_s^2 = \alpha A_s < 1$ as the consequence of $0 < \alpha \le 1$. Note that $\rho(a)$ has a lower limit, namely $\rho(a)\geq \rho_{\text{ch},0}(A_{\text{s}})^{\frac{1}{1+\alpha}}$. Note that the Chaplygin gas does not violate null energy condition $\rho+p\geq 0$ because $\rho+p=\rho-A \rho^{-\alpha} = B a^{-3(1+\alpha)} \rho^{-\alpha}$, where $\rho \ge A^{\frac{1}{1+\alpha}} > 0$ and $B = \rho_{\text{ch},0}^{1+\alpha} (1-A_s) > 0$ in the case considered. Therefore, for small values of the scale factor, the density $\rho (a)$ behaves like a dust matter $\rho(a) \approx a^{-3}$. Instead, for the large scale factor one gets  $\rho(a) = A^{1/(1+\alpha)}$, i.e. the effect of the cosmological constant. We use the idea of the Chaplygin gas \cite{Chaplygin:1904gj} because such an equation of state interpolates between the matter dominating phase and the quintessence epoch at which the $\Lambda$ is dominating. There is also an intermediate phase mimicking the Zeldovich stiff matter domination. Moreover, for $\alpha=0$ Chaplygin gas corresponds exactly to the presence of the cosmological constant (dark energy) and dust (dark and baryonic matter) \footnote{One should notice that the best fit obtained in \cite{Borowiec:2015qrp} corresponds to a small value of $\alpha=0.0194$.}. We recall that in the modified gravity framework `fluid dark energy' can be replaced by the cosmological constant ensuing from the modification of the gravitational action. Therefore, in the model under consideration, the null energy condition is not violated and the bounce is a consequence of the modification of the Einstein equations, i.e. the presence of the additional term $\gamma R^2$ in the Lagrangian, when $\gamma$ is negative.

Since the exact form of a function $f(R)$ is not known, one needs to consider some effective theories probing theoretical possibilities of this approach to gravity. Usually some a priori truncated polynomial form with respect to the scalar field and their inverse are proposed. It enables us to study how different problems of contemporary cosmology like  dark energy and dark matter issues can be solved \cite{Sotiriou:2008rp,Kamenshchik:2001cp,Allemandi:2004wn,Borowiec:2011wd,Borowiec:2015qrp,Szydlowski:2006az,Nojiri:2010wj}.
Now let us choose the simplest modification of the general relativity Lagrangian already mentioned as a simple solution of the structural equation, that is, let us consider
\begin{equation}
f(\hat{R})=\hat{R}+\gamma\hat{R}^2
\end{equation}
for which one deals with the relation
\begin{equation}
\hat{R}(T)=-T=\rho-3p.
\end{equation}
Finally, after substitution formulas (\ref{p}) and (\ref{rho}) we obtain the following $a(t)$-dependence for the Palatini scalar
\begin{equation}
\hat{R}=\left(A+Ba^{-3(1+\alpha)}\right)^{-\frac{\alpha}{1+\alpha}}\left(4A+Ba^{-3(1+\alpha)}\right).
\end{equation}
In this case the generalized Friedmann equation can be rewritten to the form of relation $H^2(a)$ as it was done in \cite{Borowiec:2015qrp}.
Following that result,
the relation $\frac{H^2}{H^2_0}$ for our model, where $H_0$ is the present value of the Hubble function, is written as
\begin{equation}
\frac{H^2}{H_0^2}=\frac{b^2}{\left(b+\frac{d}{2}\right)^2}\left(\Omega_{\gamma}\Omega_{\text{ch}}^2\frac{(K-3)(K+1)}{2b}+\Omega_{\text{ch}}+\Omega_k\right),\label{hubble}
\end{equation}
where
\begin{align}
\Omega_k &=-\frac{k}{H_0^2 a^2},\;\;K =\frac{3A_{\text{s}}}{A_{\text{s}}+(1-A_{\text{s}})a^{-3(1+\alpha)}},\;\;\Omega_{\gamma} =3\gamma H_0^2,\;\;
\Omega_{\text{ch}} =\frac{\rho_{\text{ch},0}}{3H_0^2}\left(\frac{3A_s}{K}\right)^{\frac{1}{1+\alpha}},\label{k3} \\
b &=f'(\hat{R})=1+2\Omega_{\gamma}\Omega_{\text{ch}}(K+1),\;\;d=\frac{1}{H}\frac{db}{dt}=2\Omega_{\gamma}\Omega_{\text{ch}}(3-K)[\alpha(1-K)-1]
\end{align}
and $\rho_{\text{ch},0}$ is the present value of $\rho_{\text{ch}}$, $k=-1,0,+1$ is the space curvature
and $A_{\text{s}}=\frac{A}{\rho_{\text{ch},0}^{1+\alpha}}$~\cite{Allemandi:2004wn,Borowiec:2011wd,Borowiec:2015qrp}.

This model was examined by us~\cite{Borowiec:2015qrp} where we demonstrated how the Palatini formulation modifies evolutionary scenarios with respect to the $\Lambda$CDM model so it can be considered as a natural extension of the standard cosmological model. In this paper we focus our attention on a study of the dynamics of the considered model methods provided by dynamical systems. We show that the dynamics of the model can be considered as a two-dimensional dynamical system of a Newtonian type (\cite{Szydlowski:2006az,Borowiec:2011wd}). It turns out that the phase space structure is more complicated that for the standard dynamical system because of the presence of the degenerate singularity which belongs to
type III~\cite{Nojiri:2015wsa,Odintsov:2015gba}. This singularity has an intermediate character~\cite{Herrera:2013rca} and shared evolutionary paths on two $\Lambda$CDM types of evolution (two-phases model with matter and dark energy domination epochs). The mathematical model of such dynamics is formulated with the help of notion of `sewn dynamical systems'~\cite{Bautin:1976mt}. Following this approach the full trajectories are sewn trajectories of two cuts of half-trajectories along the singularity. We have found that a weak degenerate singularity of type III appears. This is a generic feature of the dynamics in the early universe.

Barrow and Graham have introduced the concept of singular inflation \cite{Barrow:2015ora}. In this context the presence of a freeze singularity in the early evolution of the universe opens a discussion of modeling the inflation through a singularity of type III.

Let us summarize the work that we are going to present. The main goal of the paper is the investigation of the dynamics of homogeneous and isotropic cosmological model with the Lagrangian $\hat{R}+\gamma\hat{R}^2$ in the Palatini formalism. It will be demonstrated that the dynamics can be in general reduced to the form of a two-dimensional dynamical system of a
Newtonian type. This enables us to investigate the dynamics in details in the configuration space as well as the phase space where the phase portrait revealing the global dynamics can be constructed. Due to this representation of dynamics it is possible to study all evolutionary paths for all admissible initial conditions.

\section{Dynamical system approach in study of evolution of the Universe}

Our case belongs to a class of cosmological models of modified gravity whose dynamics can be reduced to the form of a two-dimensional
dynamical system in a Newtonian form~\cite{Szydlowski:2006az,Borowiec:2011wd}. Lagrangians have the form of the
one for natural mechanical systems. Therefore, the Hamiltonian $H$ has a kinetic term quadratic in momenta and the potential as a function of
state variables. The motion of the system is along the energy levels $H=E =\text{const}$. Due to this reduction the Universe dynamics can be treated as a particle of unit mass moving in the one-dimensional potential. This enables us to classify all evolutionary paths in the configuration
space and in the phase one as well. Dynamical systems of a Newtonian type are special because they describe the evolution of conservative systems like in classical mechanics. For such systems the construction of a phase portrait can be obtained directly from a functional form of the potential.

In cosmological implications a positional variable constitutes the scale factor $a(t)$ while the localization of the critical points as well
as their type are determined by a shape of the potential $V(x)$. Let us remind some commonly used terminology and properties:
\begin{enumerate}
\item A static universe is represented by a critical point of
the system $\dot{x} = y$, $\dot{y}=-\frac{\partial V}{\partial x}$ and it always lies on the $x$-axis,, that is, $y=y_0 = 0$, $x = x _0$.
\item We will say that the point ($x_0$, 0) is a critical point of a Newtonian system if that is a critical point of the function of the potential
$V(x)$, it means: $V (x) = E$, where $E= \frac{y^2}{2} + V (x)$ is total
energy of the system. Spatially flat models will admit the case $y=\dot{x}$; $E = 0$ while the ones with the spatial curvature $k\neq0$ (constant) have
$E=-\frac{k}{2}$.
\item A critical point ($x_0$ , 0) belongs to a saddle type if it is a strict local maximum of the potential $V(x)$.
\item If ($x_0$ , 0) is a strict local minimum of the analytic function $V (x)$ then one deals with
a center.
\item ($x_0$, 0) is a cusp it it is a horizontal inflection point of the $V(x)$.
\end{enumerate}

From the above it should be clear that the shape of potential function determines the critical points and its stability.
The integral of energy levels defines
the algebraic curves in the phase space ($x$, $y$) which are mimicking the evolution of the system with time. The eigenvalues of
the linearized matrix satisfy the characteristic equation of the form $\lambda^2 + \frac{\partial ^2}{\partial x ^2} V|_{x=x_ 0} =0$. They may be
real or imaginary with vanishing real parts (non-hyperbolic critical points) and hence there are another critical points beyond
the centers and saddles. The centers are structurally unstable~\cite{Perko:2001de} in opposite to saddles which represent structurally
stable critical points.

\section{Classification of the trajectories representing evolution of the model}
\subsection{Classification of the possible evolutional scenarios in the configurational space}

In order to classify all evolutionary paths in the configuration space we treat the cosmic evolution as a simple mechanical system with the
natural form of Lagrangian $L=\frac{\dot{a}^2}{2}-\tilde{V}(a)$.

From the formula (\ref{hubble}) after the substitution $H^2\equiv \left(\frac{\frac{da}{dt}}{a}\right)^2$ we obtain
that
\begin{equation}
\frac{\dot{a}^2}{2}+\tilde{V}(a)=0,\label{friedmann2}
\end{equation}
where from now on we will use dot as $\dot{} \equiv \frac{d}{d\tau}$ and $\tau$ is rescaled cosmological time, that is, $H_0 t=\tau$. The
potential $\tilde{V}$ is defined as
\begin{equation} \label{eq:tildeV}
\tilde{V}=-\frac{ a^2}{2}\frac{b^2}{\left(b+\frac{d}{2}\right)^2}\left(\Omega_{\gamma}\Omega_{\text{ch}}^2\frac{(K-3)(K+1)}{2b}+\Omega_{\text{ch}}+\Omega_k\right).
\end{equation}

As it was already mentioned, the equation (\ref{friedmann2}) has a simple mechanical interpretation of the evolutionary of the universe in term of positional variable $a(t)$. Hence the model dynamics has dynamics of particle motion of unit mass in the potential $\tilde{V}$ over the energy level.

Such an interpretation enable us to examine admissible trajectories and their classification in configuration
and phase space. All information about dynamics are coded in the geometry of the potential function.

\begin{figure}
  \centering
  \includegraphics[width=0.7\linewidth]{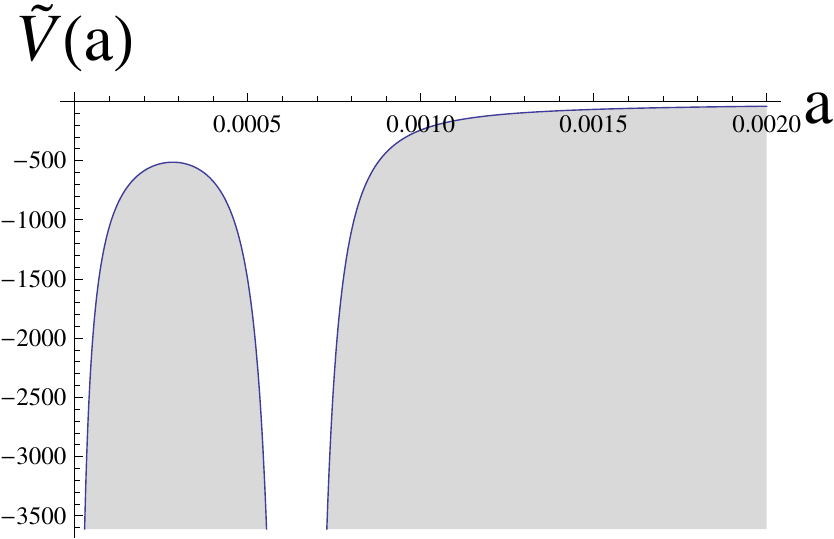}
  \caption{The diagram presents the potential $\tilde{V}(a)$ for $A_s=0.7264$, $\alpha=0.0194$ and $\Omega_{\gamma}=10^{-9}$. The shaded region represents a non-physical domain forbidden for motion of a classical system for which $\dot{a}^2\geq 0$.}
  \label{fig:3}
\end{figure}

The diagram of the potential function~(\ref{eq:tildeV}) for typical values of model parameters is presented in fig.~\ref{fig:3}.

We also define function $V(a)$ in the form
\begin{equation}
V=-\frac{ a^2}{2}\left(\Omega_{\gamma}\Omega_{\text{ch}}^2\frac{(K-3)(K+1)}{2b}+\Omega_{\text{ch}}+\Omega_k\right).
\end{equation}

\begin{figure}
  \centering
  \includegraphics[width=0.7\linewidth]{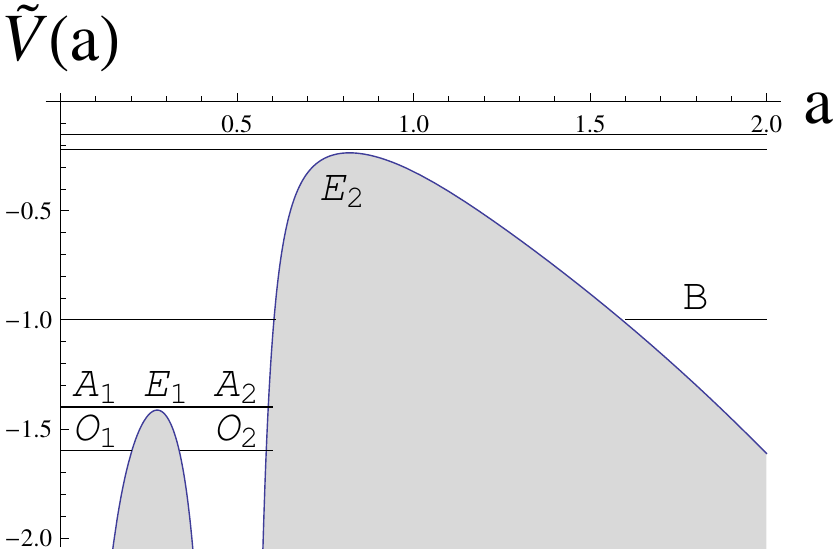}
  \caption{Diagram of the potential of dynamical system of a Newtonian type. The classification of trajectories are presented the configuration
  space. The shaded domain $E-V<0$ is forbidden for motion of classical systems.}
  \label{fig:4}
\end{figure}

The motion of system takes place in the configuration space $\{a \colon a\geq 0\}$ over the `energy' level $E=0$, i.e., the Hamiltonian is of the form $\mathcal{H}(p,a)=\frac{1}{2}p^2_a+V(a)=0\equiv E=0$. Different energy levels determine corresponding types of evolution (fig.~\ref{fig:4}).

The boundary of the domain admissible for motion is
\begin{equation}
D_{E=0}=\{a \colon V\leq 0\}.
\end{equation}
Of course it is set with the boundary
\begin{equation}
\partial D_{E=0}=\{a \colon V=0\}.
\end{equation}
Note that domain $E-V<0$ beyond this boundary is forbidden for classical motion. Let us classify all evolutionary scenarios in the configuration space:

\begin{enumerate}
\item $O_1$ --- oscillating universes with initial singularities;
\item $O_2$ --- `oscillatory solutions' without the initial and final singularity but with the freeze singularity;
\item $B$ --- bouncing solutions;
\item $E_1$, $E_2$ --- solutions representing the static Einstein universe;
\item $A_1$ --- the Einstein-de Sitter universe starting from the initial singularity and approaching asymptotically static Einstein universe;
\item $A_2$ --- a universe starting asymptotically from the Einstein universe, next it undergoes the freeze singularity and approaches to a maximum
size. After approaching this state it collapses to the Einstein solution $E_1$ through the freeze singularity;
\item $A_1$ --- expanding universe from the initial singularity toward to the Einstein universe $E_2$ with an intermediate state of the freeze singularity;
\item $EM$ --- an expanding and emerging universe from a static $E_2$ solution (Lemaitre-Eddington type of solution);
\item $I$ --- an inflectional model (the relation $a(t)$ possesses an inflection point), an expanding universe from the initial singularity undergoing the freeze type of singularity.
\end{enumerate}
The last two solutions $EM$ and $I$ lie above the maximum of the potential $\tilde{V}$.

Moreover, one deals with a singularity at which the acceleration $\ddot{a}$ is undefined because left-hand side limit of the derivative of
the potential is positive while the right-hand side limit is negative. This kind of a singularity should be treated as sewing two
singularities: one type III singularity and one type III singularity with reverse time. In other words, this special type of singularities goes
beyond the classification of four types of finite-time singularities \cite{Nojiri:2005sx,Singh:2010qa}.

\subsection{Phase portrait from the potential}
Due to Hamiltonian formulation of dynamics it is possible also to classify all evolution paths in the phase space by constructing the phase portrait of the system
\begin{align}
p &= \dot a =x \label{eq:ds1}\\
\ddot a &= \dot x= -\frac{\partial \tilde{V}(a)}{\partial a} = \frac{x^2}{ m}\frac{\partial m}{\partial a}
-m^2\frac{\partial V(a)}{\partial a} \label{eq:ds2}
\end{align}
on the phase plane $(a,x)$ with the constraint $ a'^2=-2V(a)$, i.e. $\tau-\tau_0=\int_0^a\frac{da}{\sqrt{-2V(a)}}$
and $'\equiv\frac{d}{d\sigma}=\frac{b+\frac{d}{2}}{b}\frac{d}{d\tau}$. The quantity $m=\frac{b}{b+\frac{d}{2}}$.

It would be useful to look on the dynamical problem from the point of view of sewn dynamical systems~\cite{Hrycyna:2008gk,Ellis:2015bag}. Our strategy
is following. For construction of a global phase portrait one divides dynamics in two parts, that is, one for the scale factor $a<a_{\text{fsing}}$ and
other for $a>a_{\text{fsing}}$. Such a construction divides the configuration space which is glued along the singularity.

The re-parametrization $\tau\rightarrow \sigma$ can be performed and corresponding dynamical systems assumes the following form for $a<a_{\text{fsing}}$
\begin{align}
\dot a &= x,\\
\dot x &=\frac{x^2}{m}\frac{\partial m}{\partial a}-m^2\frac{\partial V_1(a)}{\partial a},
\end{align}
where $V_1=V(-\eta (a-a_s)+1)$ with respect to the new time $\sigma$ and $\eta(a)$ notes the Heaviside function.

\begin{figure}
  \centering
  \includegraphics[width=0.7\linewidth]{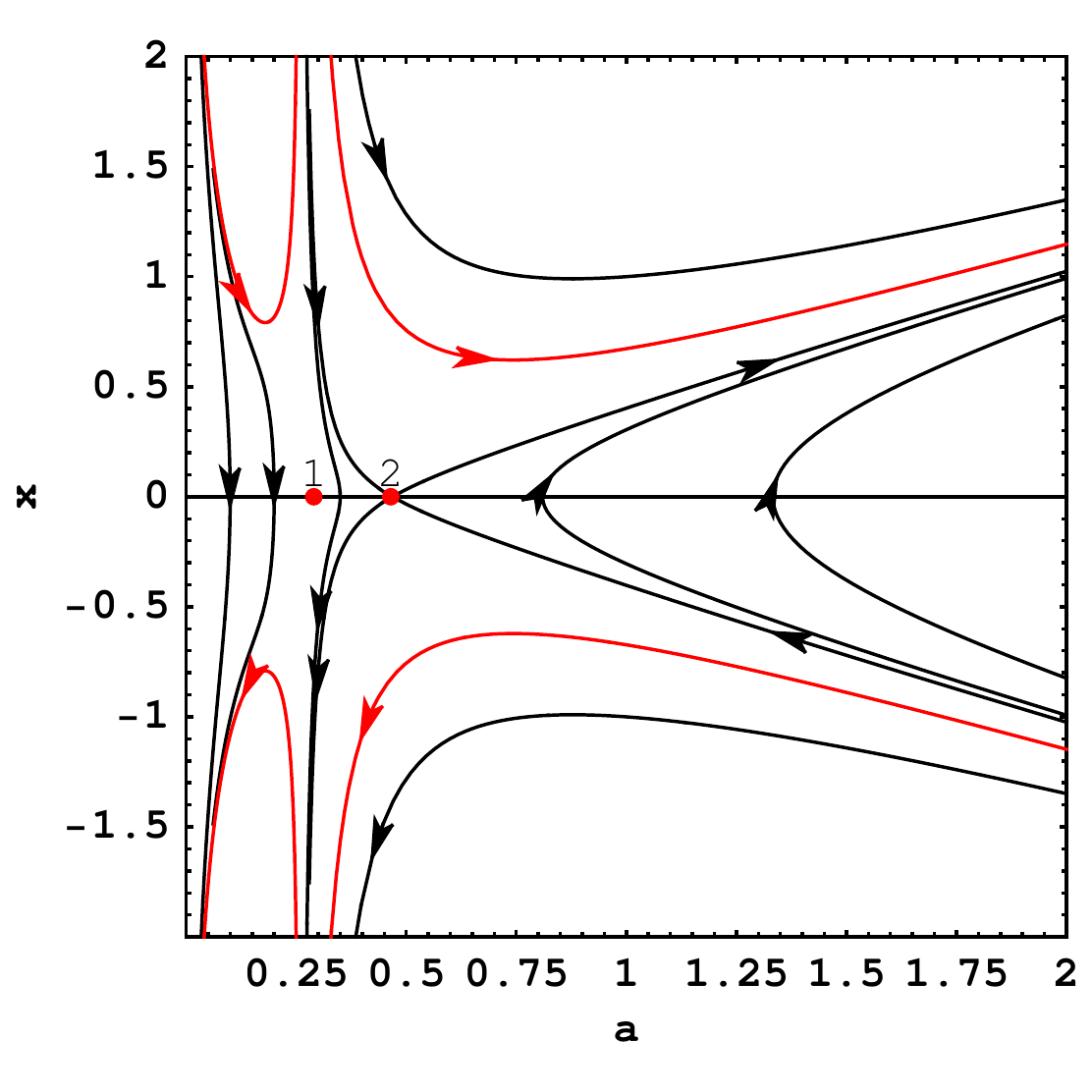}
  \caption{The diagram represents the phase portrait of system (\ref{eq:ds1}-\ref{eq:ds2}) for positive $\Omega_{\gamma}$. The red
  trajectories represent the spatially flat universe. Trajectories under the top red trajectory and below the bottom red trajectory represent
  models with negative curvature. Trajectories between the top and bottom red trajectory are ones with positive curvature. The phase
  portrait belongs to the class of the sewn dynamical systems \cite{Bautin:1976mt}. Point of sewing is located at
  infinity ($a=a_{\text{fsing}}$, $\dot{a}=\infty$). Note that all trajectories of open models are passing through the freeze singularity.
  The phase portrait possesses the reflectional symmetry $x\rightarrow -x$. Trajectories from the domain $x<0$ continue their evolution into domain $x>0$.
  Due to this symmetry one can identify the corresponding point on the line $\{ b=0 \}$ and make from the line $\{ b=0 \}$ a circle $S^1$.
  Therefore the phase space is a cylinder. The line $\{ b=0 \}$ is not shown.}
  \label{fig:5}
\end{figure}

In an analogous way for the domain configuration space $\{a\colon a>a_{\text{fsing}}\}$ we have
\begin{align}
\dot a &=x,\\
\dot x &=\frac{x^2}{m}\frac{\partial m}{\partial a}-m^2\frac{\partial V_2(a)}{\partial a}
\end{align}
where $V_2=V \eta (a-a_s)$ where $\eta$ is the Heaviside function. The phase portrait is presented
in fig.~\ref{fig:5} and belongs to the class of sewn dynamical systems. Recently such systems have been
applied to the modeling of cosmological evolution and inflation~\cite{Nojiri:2015wsa,Odintsov:2015gba,Odintsov:2015wwp}.

The classification of all possible evolutionary paths in the phase space completed our previously classification in the configuration space.

Note that trajectory represented the evolution of our Universe is located in a close neighborhood of a trajectory of the spatially flat
model, i.e. a universe is expanding and starting from the initial singularity, going through a freeze singularity and after accelerating phase
is going toward the de Sitter attractor.

\subsection{Classification of the trajectories for $\Omega_{\gamma}<0$}

For completeness let us consider the case of the spatially flat model with $\Omega_{\gamma}<0$. Because the value of $\gamma=0$ is a bifurcation
parameter (the dynamics qualitatively changes under the change of sign $\gamma$) we consider separately both these cases.

In this case one can apply a simple method of the classification of the evolutionary paths based on the consideration of a boundary of the
domain admissible for a motion in the configuration space $H^2\equiv 0$.

\begin{figure}
	\centering
	\includegraphics[width=0.7\linewidth]{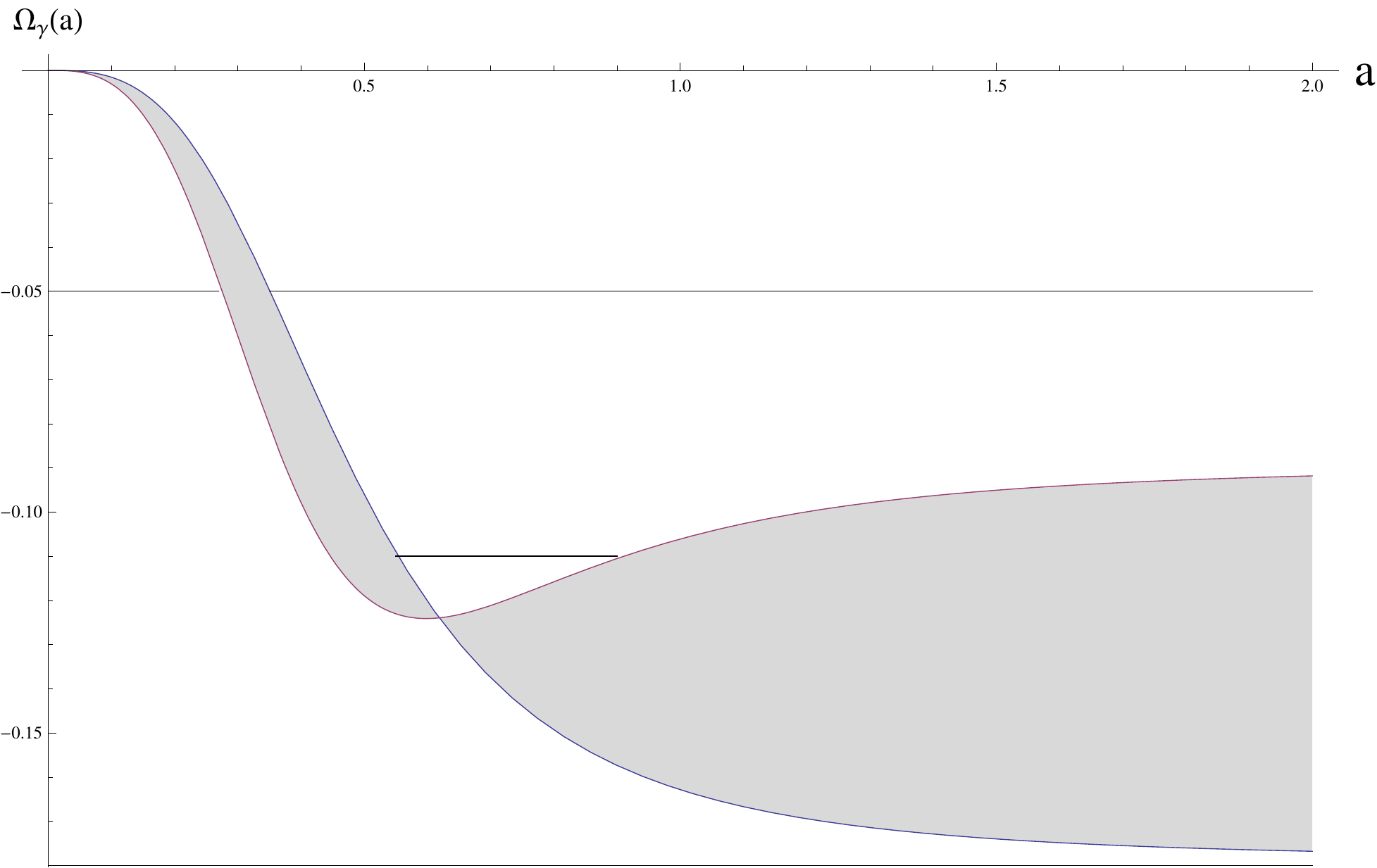}
	\caption{Diagram of $\Omega_{\gamma}(a)$ dependence used for a classification of all evolutional paths of the spatially flat models
	with a negative $\gamma$. The type of evolution we obtain after consideration of levels $\Omega=\Omega_{\gamma}$. We can simply discover
	oscillating models without the initial and final singularity, oscillating models with the initial and final singularity, models evolving
	to infinity with the initial singularity and bouncing models. From this analysis we conclude that for the physical case of sufficiently small
	values of $\Omega_{\gamma}$ all models possess bounce instead of the initial singularity. The blue line represents
	$ b=0 $ (equation (\ref{gamma})). Above the blue line is the region for $ b>0 $ and this region corresponds to the physical domain.
	The region below the blue line is for $ b<0 $ and represents the non-physical domain. The red line represents function (\ref{gamma2}). Between
	the red line and the blue line $ H^2<0 $ (light gray domain). The gray domain also represents non-physical region.}
	\label{fig:7}
\end{figure}

From (\ref{hubble}) the equation of the boundary curve assumes the following form
\begin{equation}
\left(\frac{1+2\Omega_{\gamma}\Omega_{\text{ch}}(K+1)}{1+\Omega_{\gamma}\Omega_{\text{ch}}(\alpha K^2+(3-4\alpha)K-1+3\alpha)}\right)^2 \left(\Omega_{\gamma}\Omega_{\text{ch}}\frac{(K-3)(K+1)}{2(1+2\Omega_{\gamma}\Omega_{\text{ch}}(K+1))}+1\right)\equiv 0,\label{boundary}
\end{equation}
where $\Omega_{\text{ch}}=\Omega_{\text{ch,0}}\left(\frac{3A_{\textbf{s}}}{K}\right)^{\frac{1}{1+\alpha}}$.
This idea of classification comes from the celestial mechanics where one may classify solutions through the analysis of curve of
zero velocity, i.e., levels of constant of $\Omega_{\gamma}$. Finally
\begin{equation}
\Omega_{\gamma}(a)=-\frac{1}{\Omega_{\text{ch},0}\left(A_{\text{s}}+(1-A_{\text{s}})a^{-3(1+\alpha)}\right)^{\frac{1}{1+\alpha}}\left(\frac{3A_{\text{s}}}{A_{\text{s}}+(1-A_{\text{s}})a^{-3(1+\alpha)}}+1\right)},\label{gamma}
\end{equation}
or
\begin{equation}
\Omega_{\gamma}(a)=-\frac{2}{\Omega_{\text{ch},0}\left(A_{\text{s}}+(1-A_{\text{s}})a^{-3(1+\alpha)}\right)^{\frac{1}{1+\alpha}}\left(\frac{3A_{\text{s}}}{A_{\text{s}}+(1-A_{\text{s}})a^{-3(1+\alpha)}}+1\right)^2},\label{gamma2}
\end{equation}
The diagram of $\Omega_{\gamma}(a)$ function derived from (\ref{boundary}) is shown in fig.~\ref{fig:7}.

The functions (\ref{gamma}) and (\ref{gamma2}) are always negative and their approximations for large and small value of the scale factor are, respectively:
\begin{equation}
\Omega_{\gamma}(a)=-\frac{a^{3}}{\Omega_{\text{ch,0}}(1-A_{\text{s}})^{\frac{1}{1+\alpha}}\left(\frac{3A_{\text{s}}a^{3(1+\alpha)}}{(1-A_{\text{s}})}+1\right)}
\end{equation}
or
\begin{equation}
\Omega_{\gamma}(a)=-\frac{2a^{3}}{\Omega_{\text{ch,0}}(1-A_{\text{s}})^{\frac{1}{1+\alpha}}\left(\frac{3A_{\text{s}}a^{3(1+\alpha)}}{(1-A_{\text{s}})}+1\right)^2}\text{ if } a\ll 1
\end{equation}
and
\begin{equation}
\Omega_{\gamma}(a)=-\frac{1}{4\Omega_{\text{ch},0}A_{\text{s}}^{\frac{1}{1+\alpha}}}
\end{equation}
or
\begin{equation}
\Omega_{\gamma}(a)=-\frac{1}{8\Omega_{\text{ch},0}A_{\text{s}}^{\frac{1}{1+\alpha}}}\text{ if } a\gg 1.
\end{equation}
From asymptotes for the small scale factor the effects of curvature and non-zero $\Omega_\gamma$ are negligible. The dynamics of the model
is equivalent to the $\Lambda$CDM dynamics. This means that matter effect dominates. From asymptotes for the large scale factor we
obtain the effects of both matter and curvature are negligible. The dynamics corresponds to the accelerating phase caused by dark energy.

Now on, let us consider levels of $\Omega_{\gamma}=\text{const}\leq 0$. In consequence we obtain two types of possible evolutionary paths

O --- oscillating models without the initial and final singularities;

B --- models with a bounce instead of an initial  singularity like in the case of $\Omega_{\gamma}>0$.

\noindent Note that class of bouncing models as well as oscillating ones is generic.

\begin{figure}
	\centering
	\includegraphics[width=0.7\linewidth]{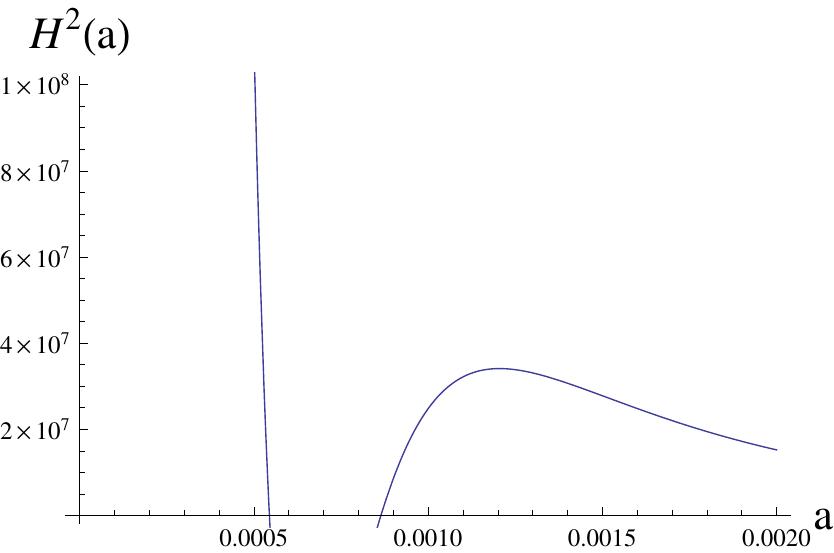}
	\caption{Diagram of $H^2(a)$ relation for the spatially flat model with a negative value of $\gamma$ parameter. Note the existence of domain forbidden for classical motion.}
	\label{fig:8}
\end{figure}

In fig.~\ref{fig:8} we plot the diagram of the $H^2 (a)$ relation. It demonstrates that the admissible domain for the motion of
classical trajectories is the union of two separated and disjoint domains occupied by the trajectories. In the domain situated on the
left in which $H^2>0$ the motion is bounded by a Big-Bang singularity: $a=0$, $H=\infty$, and a maximum value of the scale factor
on the $\Omega_{k,0}$.

\begin{figure}
	\centering
	\includegraphics[width=0.7\linewidth]{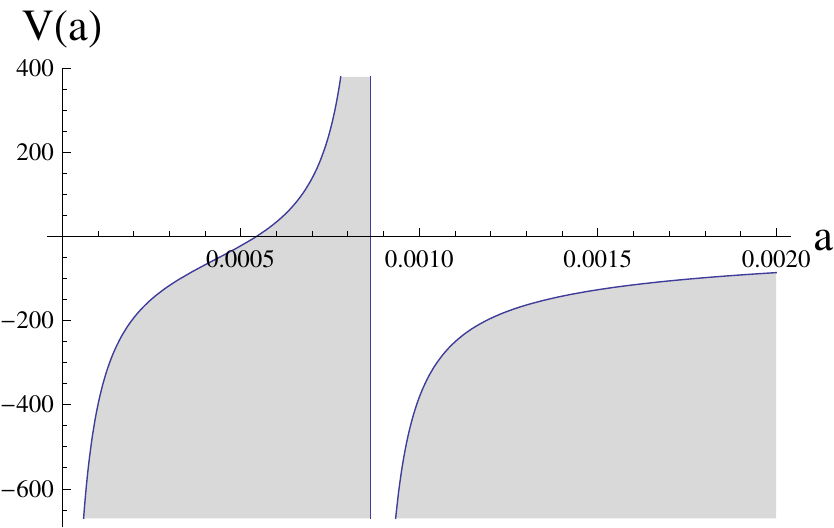}
	\caption{Diagram of the potential function $V(a)$ for the system (\ref{ds})-(\ref{ds2}). Different evolutionary paths of the
	system model are obtained by consideration energy levels $E=\frac{1}{2}\Omega_{\text{k,0}}$. The shaded regions are forbidden
	for classical motion. The vertical line has the equation $b=0$ ($a=a_{\text{sud.sing}}$).}
	\label{fig:13}
\end{figure}

In the domain $\{a\colon a>a_{\text{min}}\}$ of the configuration space there is the bouncing type of the Universe evolution.
Of course the value $a_{\text{min}}$ depends on the initial conditions, i.e., $\Omega_{k,0}$. This relation is illustrated in the
diagram of function $V(a)$ shown in fig.~\ref{fig:13}. Taking different energy levels depending on the value
of $\Omega_{k,0}$ one gets different evolutionary scenarios. In the domain of the configuration space $\{a\colon a<a_{\text{max}}\}$ we have
an oscillating solution with the big-bang which after reaching the maximum size will re-collapse to the second singularity.

In turn in the domain $\{a\colon a>a_{\text{sing}}\colon b(a_{\text{sing}})=0\}$ there are two types of trajectories:
one representing evolution of the oscillating closed models without initial and final singularities while the second one for flat $(\Omega_{k,0}=0)$,
closed $(\Omega_{k,0}>0)$ and open $(\Omega_{k,0}<0)$ models with the bounce.

There is a vertical line, which separates two disjoint domains of the configuration space. The equation $a=a_{\text{sing}}$ can be
simply obtained from the definition of $b$ function, namely
\begin{equation}
1=-2\Omega_{\gamma}\Omega_{\text{ch,0}}\left(3A_{\text{s}}\right)^{\frac{1}{1+\alpha}}K^{\frac{-1}{1+\alpha}}(K+1).
\end{equation}
In the special case of the Chaplygin gas $(\alpha=1)$ we obtain
\begin{equation}
\left(-2\Omega_{\gamma}\Omega_{\text{ch,0}}\left(3A_{\text{s}}\right)^{\frac{1}{2}}\right)^{-1}=K^{\frac{-1}{2}}(K+1),
\end{equation}
and hence one deals with the algebraic equation of second order
\begin{equation}
K^2+(2-\beta^2)K+1=0,\label{ksquare}
\end{equation}
where $\beta^2=\left(12\Omega_{\gamma}^2\Omega_{\text{ch,0}}^2A_{\text{s}}\right)^{-1}$.

Because $K\in[0,3)$, the above equation has one solution which is
\begin{equation}
K_{\text{sing}}=\frac{\beta^2-2-\sqrt{\left(2-\beta^2\right)^2-4}}{2}
\end{equation}
 or, in the terms of the scale factor $a_{\text{sing}}$
\begin{equation}
a_{\text{sing}}=\left(\frac{1}{1-A_{\text{s}}}\left(\frac{6A_{\text{s}}}{\beta^2-2-\sqrt{\left(2-\beta^2\right)^2-4}}-A_{\text{s}}\right)\right)^{\frac{-1}{3(1+\alpha)}}.
\end{equation}
Equation (\ref{ksquare}) has a real solution for $K\in[0,3)$ if the parameter $\beta^2>4$.

\begin{figure}
	\centering
	\includegraphics[width=0.7\linewidth]{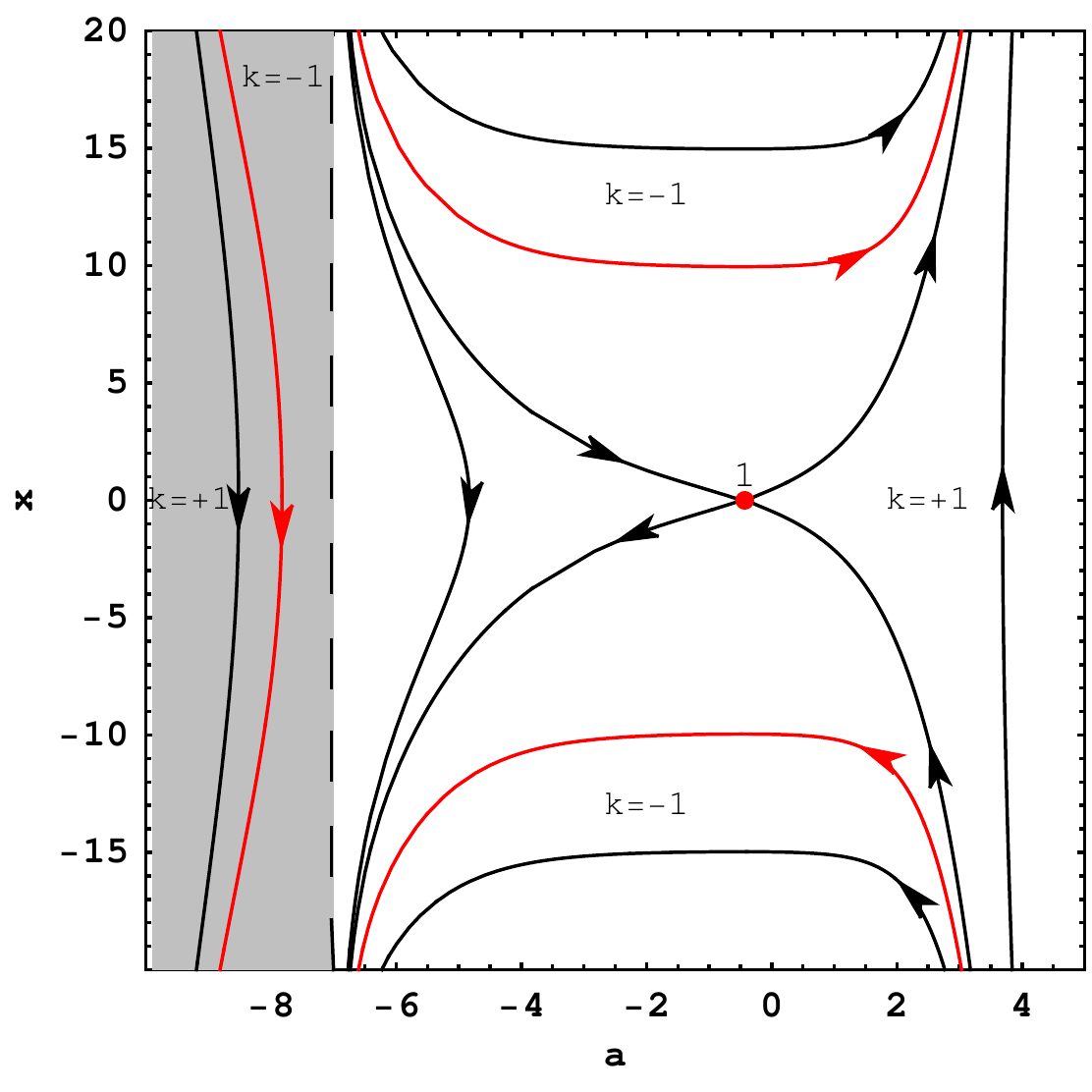}
	\caption{The phase portrait of the model with the negative $\gamma$ for the flat, closed and open models for
	the system (\ref{ds})-(\ref{ds2}). The scale factor $a$ is measured in the logarithmic
	scale. In the generic case trajectories start from the sudden singularity and go toward the de
	Sitter model. The trajectory of the flat model separates closed and open ones.  Let us concentrate on the trajectory of the flat model
	(red line). The universe collapse toward the sudden singularity as $\dot{a}$ goes
	to zero. Therefore $H$ approaches zero. Because the phase portrait possesses the reflectional symmetry $\dot{a}\rightarrow-\dot{a}$ this
	singularity is the bounce. Due to this symmetry one can identify the
	corresponding point on the $b$-line and make from the $b$-line a circle $S^1$. Therefore the phase is a cylinder.
	From the physical point of view singularities at the infinity (in $\sigma$ time) should be sewn
	because they represent the same physical state. Finally the red trajectory represents a bounce type solution with glued two sudden singularities
	in the past and in the future. Note that this type of behavior is generic for the class of all models with the curvature. The shaded region
	is occupied by trajectories with $b<0$ and this region can be removed if we postulate that $f'(R)>0$. The
	dashed line is a line of singularity $b=0$.}
	\label{fig:9}
\end{figure}

On the line $\{b=0\}$ there is a singularity point (see fig.~\ref{fig:9}), that is, $a=a_{\text{sing}},\text{ }a'_{\text{sing}}=\infty$. If we go
back to the original time then $\dot{a}=0$. It is a singularity of type II called a sudden singularity at which $a$, $\rho_{\text{eff}}$, and
Hubble parameter remain finite ($\dot{H}$ diverges). They are past singularities like a big-demarrage arising in models with the
generalized Chaplygin gas.

\begin{figure}
	\centering
	\includegraphics[width=0.7\linewidth]{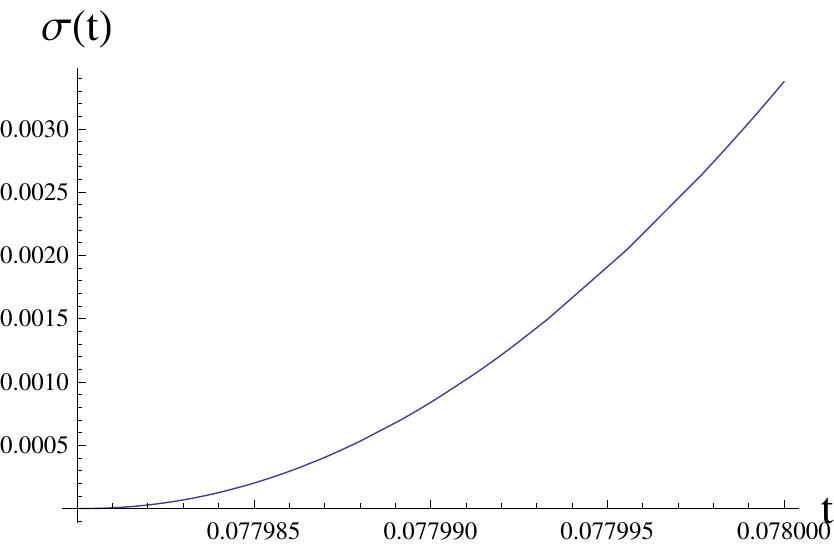}
	\caption{Diagram of the $\sigma=\sigma(t)$ time parametrization introduced for the dynamical system for the
	negative $\Omega_{\gamma}$ The new parameter $\sigma(t)$ is a monotonic function of the original cosmological time $t$. Because of the
	singularity on the line $\{b=0\}$ this parametrization is not a diffeomorphism. We assume that $8\pi G=1$ and we
	chose $\frac{\text{s Mpc}}{\text{100 km}}$ as a unit of time $t$.}
	\label{fig:14}
\end{figure}

\begin{figure}
	\centering
	\includegraphics[width=0.7\linewidth]{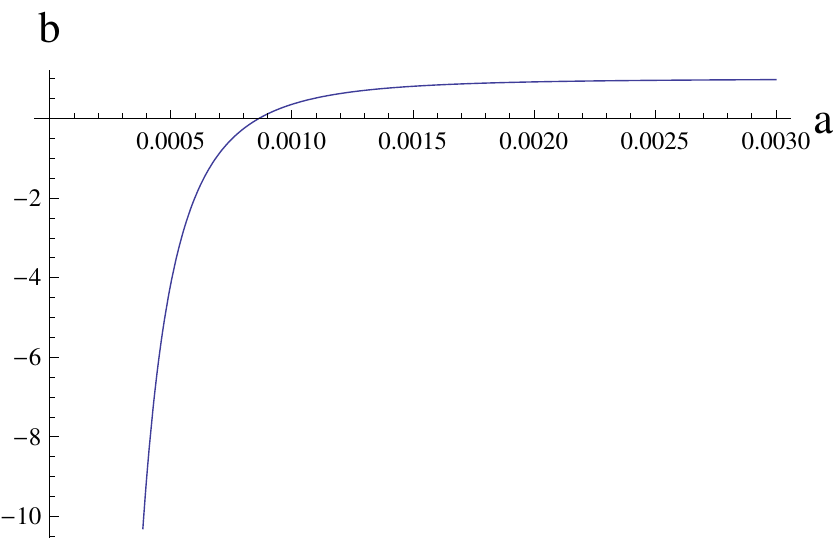}
	\caption{Diagram of the function $b(a)$. It illustrates the sign changing for the negative $\Omega_{\gamma}$.}
	\label{fig:11}
\end{figure}

\begin{figure}
	\centering
	\includegraphics[width=0.7\linewidth]{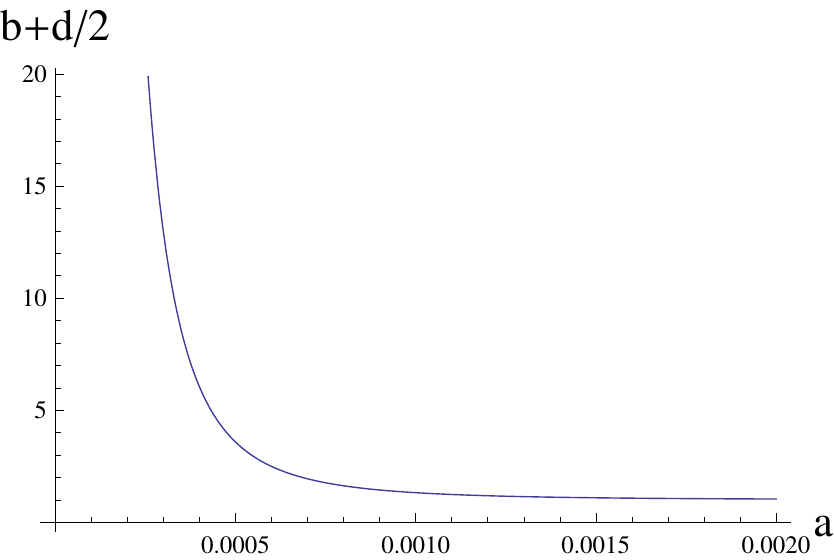}
	\caption{Diagram of function $b+d/2$ versus the scale factor for the negative $\Omega_{\gamma}$. For $a$ going
	to zero this function blows up to infinity while for $a \gg 1$ the function goes to 1.}
	\label{fig:10}
\end{figure}

Figure~\ref{fig:9} is the phase portrait of the original system under re-parametrization of
time $t\rightarrow\sigma \colon d\sigma=\frac{|b|}{|b+d/2|}dt$. The function $\sigma=\sigma(t)$ is drawn in the Fig.~\ref{fig:14}. For this
case the dynamical system is expressed by
\begin{align}
\frac{da}{d\sigma} &= x \label{ds}\\
\frac{dx}{d\sigma} &= -\frac{\partial V(a)}{\partial a}. \label{ds2}
\end{align}
Of course this re-parametrization is singular on the line $\{b=0\}$. One should notice that there is no inverse
transformation $t=t(\sigma)$. Figure~\ref{fig:11} illustrates that the function $b(a)$ changes the sign if it passes through the
zero on the $a$-axis. On the other hand the function $(b+d/2)$ is not singular (see fig.~\ref{fig:10}). Since
the re-parametrization is a non-smooth function when $b$ changes the sign, the corresponding dynamical system possesses
a discontinuity point on the line $\{b=0\}$.

All properties of the model dynamics under consideration are summarized in the phase portrait which is a picture of
the global dynamics (see fig.~\ref{fig:9}). The phase space is a union of disjoint
domains $A=\{a\colon a<a_{\text{sud.sing}}\}$ and $B=\{a\colon a>a_{\text{sud.sing}}\}$. In the both regions
trajectories representing flat ($k=0$), open ($k=-1$) and closed (k=+1) models appeared. In any case trajectory of the flat model separates
models with $k=+1$ from those with $k=-1$. In the region $A$ all solutions are oscillating and possess the initial and final singularity. In
the domain $B$ we obtain dynamics equivalent to the dynamics of $\Lambda$CDM model but in the enlarged phase space $\{(\dot{a},a)\colon a\gg 0\}$. In
the domain of $B$ the curvature effect is negligible near the singularity of the finite scale factor. The typical trajectory starts
this critical point (representing a sudden singularity) and evolves toward the Sitter universe where effects of the curvature are also negligible.
On the phase portrait there is also the critical point of the saddle type. This critical point corresponds to the maximum
of the potential $V=V(a)$. That is, it is a decreasing function of the argument so the universe is decelerating one. Therefore the
trajectories on the right-hand side of the saddle point represents accelerating models.

Finally, from the comparison of $\Lambda$CDM dynamics with our model it is seen that the initial singularity is replaced by the sudden singularity in the past
like the demarrage singularities appeared in the cosmological models with the Chaplygin gas.

\begin{figure}
	\centering
	\includegraphics[width=0.7\linewidth]{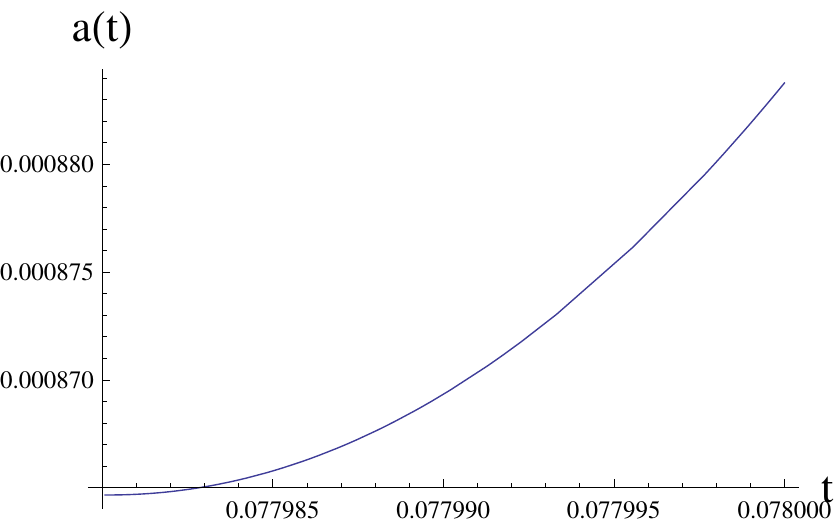}
	\caption{Diagram of the scale factor $a(t)$ for the flat model with the negative $\gamma$. This type of behavior well approximates
	the behavior of the closed and open models near the sudden singularity: the scale factor, $\rho_{\text{eff}}$, as well as the
	Hubble parameter are finite and $\dot{H}$ diverge. Note that for the small values of a time derivative $\dot{a}$ goes to zero. Therefore,
	 $H$ is asymptotically zero. We assume that $8\pi G=1$ and the unit of time $t$ is $\frac{\text{s Mpc}}{100\text{ km}}$.}
	\label{fig:12}
\end{figure}

\begin{figure}
	\centering
	\includegraphics[width=0.7\linewidth]{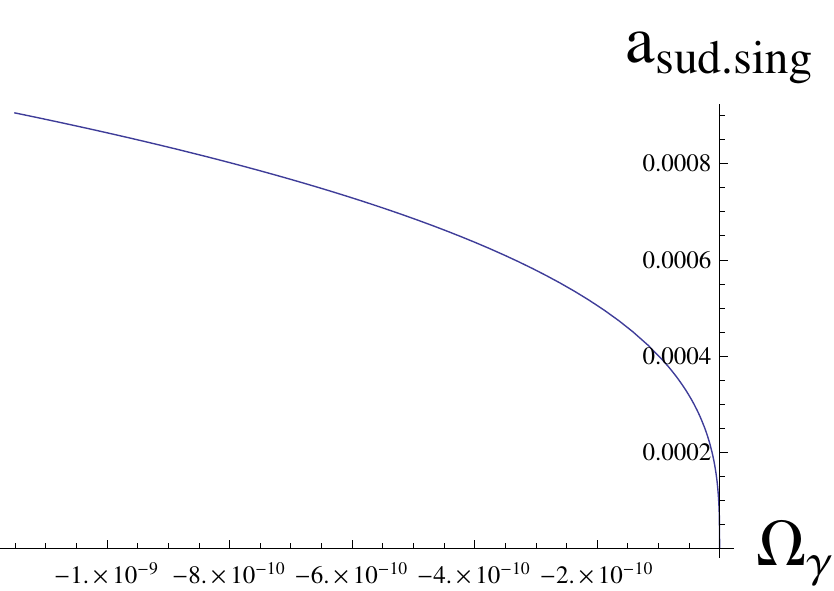}
	\caption{Diagram of the relation between positive $\Omega_{\gamma}$ and $a_{\text{sud.sing}}$ obtained for $A_s=0.6908$ and $\alpha=0.0373$. We see
		that this relation is a monotonic function. If $\Omega_{\gamma}\longrightarrow 0$ then $a_{\text{fsing}}\longrightarrow 0$.}
	\label{fig:15}
\end{figure}

In fig.~\ref{fig:12} it is shown the diagram of the scale factor for the flat models which the generic description of the evolution near sudden
singularity. On the phase portrait this singularity lies on the circle at infinity $a'^2+a^2=\infty$, where $a=a_{\text{sud.sing.}}$. If we return
to the original cosmological time $t$ then $\dot{a}$ at this critical point at the infinity is corresponding $\dot{a}=0$. In the consequence the
Hubble parameter is finite. For the FLRW model with initial singularity this follows divergence of conformal time $\int\frac{dt}{a(t)}$
because $\int\frac{dt}{a(t)}>\frac{1}{\text{const}}\int\frac{dt}{t}$ diverge as $t\rightarrow 0$. The diagram of the relation between $a_{\text{sud.sing.}}(\Omega_{\gamma})$ is presented in fig.~\ref{fig:15}.

\section{Degenerate singularity of type III as a model of endogenous intermediate inflation}

From the formula (\ref{hubble}) it is possible to detected singularity of type III, called also a freeze singularity. Our analysis shows that this type singularity is a generic property of the early evolution of the universe.

If we consider singularities in FLRW models, which is filled of perfect fluid with effective energy density $\rho_{\text{eff}}$ and
pressure $p_{\text{eff}}$ then all singularities can be classified on the four groups \cite{Singh:2010qa}. The first class is a Big Rip (type I)
singularity, where the energy density, pressure, and the scale factor diverge. Sudden singularity (type II) happens when
the scale factor and effective energy density are finite values but pressure diverges.
Big Freeze singularity (type III) is observed when effective energy
density and pressure diverge at a finite value of the scale factor while Big Brake (type IV) for the finite scale factor, effective energy density,
and pressure with a divergence in the time derivative of the pressure or change of energy density rate. In that context our singularities belongs
to the type III.

In our approach effective energy density and pressure (as well as the coefficient of the equation of state $w_{\text{eff}}$) can be simply
expressed in terms of the potential, namely
\begin{equation}
\rho_{\text{eff}}= -\frac{6V}{a^2},
\end{equation}
\begin{equation}
p_{\text{eff}}=-\rho_{\text{eff}}-\frac{1}{3}\frac{d(\rho_{\text{eff}})}{d(\ln a)},
\end{equation}
\begin{equation}
w_{\text{eff}}==-1-\frac{1}{3}\frac{d(\ln \rho_{\text{eff}})}{d(\ln a)}.
\end{equation}
In our case potential as well as the effective pressure diverges. This singularity is the singularity of acceleration as the derivative
of the potential goes to plus infinity on the left of this point while on the right side one has minus infinity. At this
singularity $da/dt$ also diverge while the scale factor is finite. In the diagram of the scale factor as a function of time one can observe how
the function $a(t)$ changes an inflection along the vertical line $t=t_{\text{fsing}}$.

Such types of singularities appear in the context of Loop Quantum Cosmology \cite{BouhmadiLopez:2006fu,Singh:2010qa,Kiefer:2010zz,Odintsov:2014gea} where are called a hyper-inflation state \cite{Hrycyna:2008yu}. There exists a close relation between Palatini gravity and an effective action of Loop Quantum Gravity that reproduces the dynamics \cite{Olmo:2008nf,Olmo:2008pv} of considered $f(R)$ models of polytropic spheres in Palatini formalism $f(R)=R\pm\lambda R^2$ ($\lambda$ has order of squared Planck length). The freeze type of singularity in the model under consideration is a solution of the algebraic equation
\begin{equation}
(2b+d)^2 =0 \Longrightarrow f(K,\alpha,A_{\text{s}},\Omega_{\gamma})=0
\end{equation}
or
\begin{equation}
\alpha K^2-3(1+\alpha)K-\frac{K^{\frac{1}{1+\alpha}}}{\Omega_{\gamma}\Omega_{\text{ch,0}}\left(3A_{\text{s}}\right)^{\frac{1}{1+\alpha}}}+1=0,\label{k}
\end{equation}
where $K\in [0,\text{ }3)$.

If $\alpha=0$ then equation~(\ref{k}) simplifies to
\begin{equation}
-3K-\frac{K}{\Omega_{\gamma}\Omega_{\text{ch,0}}\left(3A_{\text{s}}\right)}+1=0.
\end{equation}
The solution of the above equation is
\begin{equation}
K_{\text{sing}}=\frac{1}{3+\frac{1}{3\Omega_{\gamma}\Omega_{\text{ch},0}A_{\text{s}}}}.\label{k2}
\end{equation}
From equation (\ref{k3}) and (\ref{k2}) one finds an expression for a value of the scale factor for the degenerated freeze singularity
\begin{equation}
a_{\text{sing}}=\left(\frac{1-A_{\text{s}}}{8A_{\text{s}}+\frac{1}{\Omega_{\gamma}\Omega_{\text{ch,0}}}}\right)^\frac{1}{{3}},
\end{equation}
or in the term of redshift
\begin{equation}
z_{\text{sing}}=\left(\frac{8A_{\text{s}}+\frac{1}{\Omega_{\gamma}\Omega_{\text{ch},0}}}{1-A_{\text{s}}}\right)^\frac{1}{{3}}-1.
\end{equation}

\begin{figure}
  \centering
  \includegraphics[width=0.7\linewidth]{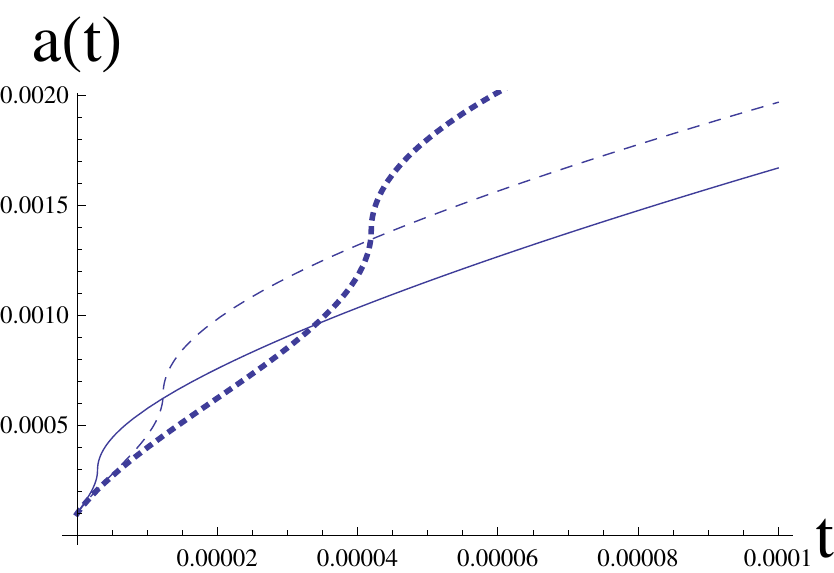}
  \caption{The diagram represents function $a(t)$ for positive $\Omega_{\gamma}$. For the scale factor of the freeze singularity,
  the function $a(t)$ has a vertical inflection point. The continuous line is for $\Omega_{\gamma}=10^{-10}$, the dashed line is
  for $\Omega_{\gamma}=10^{-9}$ and the dotted line is for $\Omega_{\gamma}=10^{-8}$. Is is assumed that $A_{\text{s}}=0.7264$ and $\alpha=0.0194$. We
  assume that $8\pi G=1$ and we chose $\frac{\text{s Mpc}}{\text{100 km}}$ as a unit of time $t$.}
  \label{fig:6}
\end{figure}

This singularity, which is the horizontal inflection singularity point $a=a_{\text{fsing}}$, corresponds to the diagram of the scale
factor of $t$ (fig.~\ref{fig:6}). Note that at this singularity the potential $\tilde{V}$ diverges.

\begin{figure}
  \centering
  \includegraphics[width=0.7\linewidth]{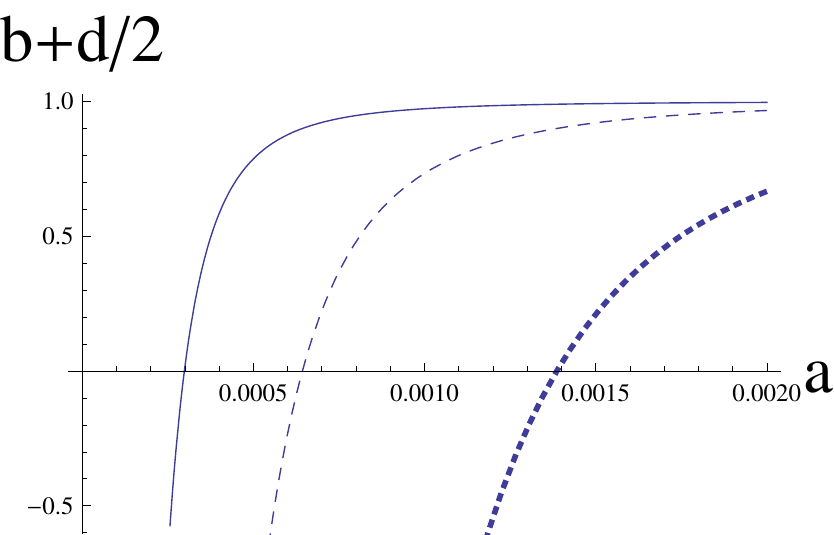}
  \caption{The diagram represents function $b(a)+d(a)/2$ for different values of the positive $\Omega_{\gamma}$ and shows that it is growing
  function of scale factor. Zero of this function represents a value of the scale factor for the freeze singularity $a_{\text{fsing}}$. The continuous
  line is for $\Omega_{\gamma}=10^{-10}$, the dashed line for $\Omega_{\gamma}=10^{-9}$ and the dotted line is for $\Omega_{\gamma}=10^{-8}$. Is is
  assumed that $A_{\text{s}}=0.7264$ and $\alpha=0.0194$. One observes a single isolated zero of function corresponding to the
  single singularity.}
  \label{fig:1}
\end{figure}

\begin{figure}
  \centering
  \includegraphics[width=0.7\linewidth]{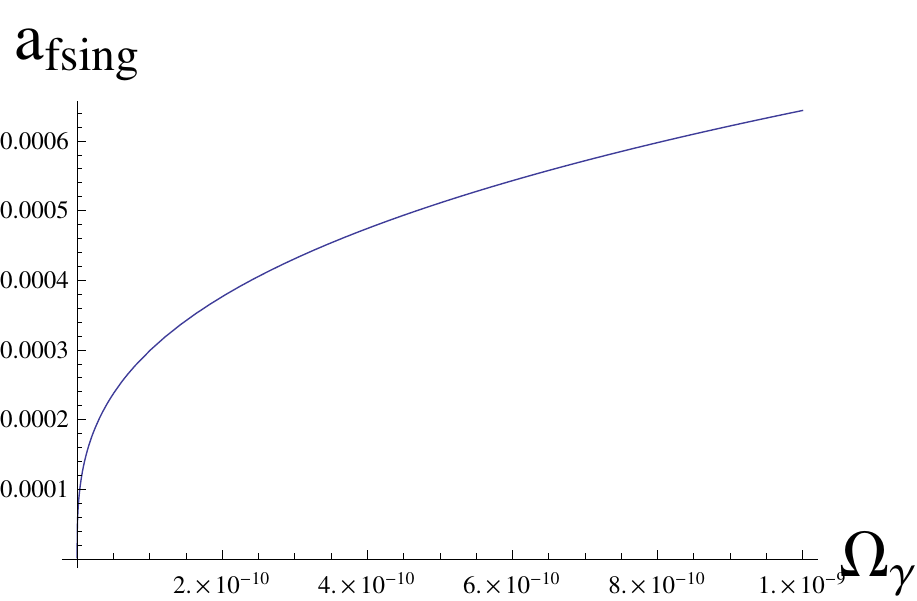}
  \caption{The diagram shows the relation between positive $\Omega_\gamma$ and $a_{\text{fs}}$ obtained for $A_s=0.7264$ and $\alpha=0.0194$. We see
  that this relation is a monotonic function. If $\Omega_\gamma\longrightarrow 0$ then $a_{\text{fsing}}\longrightarrow 0$.}
  \label{fig:2}
\end{figure}

For the special case of the Chaplygin gas ($\alpha=0$) the exact formulas or value of $a_{\text{fsing}}$ can be obtained~\cite{Borowiec:2015qrp}.
The diagram of function $b+\frac{d}{2}=f(A_{\text{s}},\Omega_{\text{ch},0},\Omega_{\gamma},\alpha)$ is presented in fig.~\ref{fig:1}.
From our numerical analysis (see fig.~\ref{fig:2}) we have obtained that a single freeze type singularity is generic property of dynamics for the broad range of model parameters ($A_{\text{s}}$, $\alpha$, $\Omega_{\text{ch},0}$, $\Omega_{\gamma}$).

\section{Singularities and astronomical observations}

In this section, we discuss the status of singularities appearing in the model under the consideration. If we compare this model with the $\Lambda$CDM model which formally can be obtained after putting $\alpha =0$ then one can conclude that the latter inherits these singularities.

Basing on the estimation of the model parameters performed in our previous paper \cite{Borowiec:2015qrp} one can calculate the value of the scale factor (or redshift) corresponding to this event in the history of the universe when singularities appear. This value depends on the value of density parameters $\Omega_\gamma$, $A_\text{s}$, $\alpha$, $H_0$. From numerical simulations we obtain that this value is only sensitive on $\Omega_\gamma$ parameter and the dependence on the parameter $\alpha$ is very weak. This means that corresponding values of redshifts obtained for these singularities in the case $\alpha \in (0,1)$ do not differ. In figures~\ref{fig:15}, \ref{fig:2} we illustrate how values of redshifts (the scale factor marked in the figures) depend on the crucial value of density parameter $\Omega_\gamma$. While for positive $\Omega_\gamma$ this function is a growing function of the scale factor, but for negative values of $\Omega_\gamma$ it is a decreasing function of the scale factor. In both cases it is a monotonic relation. Of course for the case of $\Omega_\gamma =0$ the correspondence with the $\Lambda$CDM model is achieved and both sudden and freeze singularities vanish. Therefore the presence of singularities of the model under consideration should be treated as a property which is strictly related with the Palatini formalism.

For the best fitted values of the model parameters obtained in our previous paper \cite{Borowiec:2015qrp} one could estimate simply the value of redshift corresponding of the sudden singularity at the bounce. This value is 1103.67. In the case of the positive $\Omega_\gamma$,  this parameter lies on boundary which unable us formulation of an analogous conclusion.

In the monograph by Capozziello and Faraoni \cite[p.~73]{Capozziello:2011beg}, section 3.4.2, some problems with the Palatini formalism were addressed. The authors remarked that the $f(\hat R)$ gravity suffers from two serious problems. The first problem is the presence of curvature singularities at the surface of stars and the second one is incompatibility with the Standard Model of particle physics. Let us consider our model in the context of singularities. Our remark in the context of cosmology is that the Palatini formalism rather generates singularities than suffers from the singularities like in the case of stars.

Capozziello and Faraoni also noted that different formulation of gravity can give rise to new physical effects. If we assume that freeze singularities can appear before the recombination epoch and sudden singularities being before the nucleosynthesis then such requirements should be treated as minimal conditions which guarantee that physics does not change in comparison to the $\Lambda$CDM model. In consequence if we assume that the nucleosynthesis epoch was for redshift $z=3\times 10^{8}$ (see e.g. \cite{Iocco:2008va}) then the value of $\Omega_\gamma$ parameter should belong to the interval ($-10^{-25}, 0$) for the case the negative $\Omega_\gamma$ parameter. If we analyze the likelihood functions for values of model parameters \cite{Borowiec:2015qrp} for the negative $\Omega_\gamma$ parameter then we get that the probability of appearing the sudden singularity after the nucleosynthesis epoch is $1-10^{-16}$ while the probability of appearing one before the nucleosynthesis epoch is $10^{-16}$. In consequence one can reject the case with the negative values of $\Omega_\gamma$ even it is favored by the data. If the recombination epoch takes place after the freeze singularity then it is required that the values of the positive $\Omega_\gamma$ parameter belong to the interval ($0, 10^{-9}$).

\section{Conclusions}

We have classified all evolutionary paths of the cosmological model $f(\hat{R})=\hat{R}+\gamma\hat{R}^2$ the Palatini formulation due to reduction of model dynamics to the dynamical system of
a Newtonian type. We have found a phase space structure of dynamics organized through the two saddle points representing the static Einstein universe and center.

Moreover, the localization of the freeze singularity during the cosmic evolution of the early universe was presented. We investigated in details how this
localization depends on density parameters for contribution originating from the presence of $R^2$ term in the Lagrangian.
The model possesses four phases: a first matter dominating deceleration phase, an intermediate phase of inflation, a second matter
dominating deceleration phase, and a late accelerating phase of evolution of the current universe. This evolutionary scenarios becomes in agreement
with the $\Lambda$CDM model only if the freeze singularity is shifted to the matter dominating initial singularity.

While in our case acceleration (and in consequence pressure) is undefined (note that all four types assume that
acceleration is a well defined function for which we calculate the limits). It is similar to a singularity of type III (finite scale
factor singularity) at which the scale factor remains finite, but both $\rho$ and $p$ diverge (as well as the Hubble parameter $H$). A particular
example of such a singularity is a `big-freeze' singularity (both in the past and the future).
They are characterized by the generalized Chaplygin equation of state \cite{Kiefer:2010zz}. In the model under consideration one deals with the situation
in which freeze singularities in the past and in the future are glued. This type of degeneration is beyond the standard classification. We
will call this type of weak singularity a degenerate freeze singularity. Note that trajectories in the phase space pass through this singularity and continue their evolutions.

We would like to draw the reader's attention to the conclusion that the dynamics presented in terms of dynamical systems of a Newtonian
type enable us to classify all evolutionary paths. That it, the dynamics is reduced to the 2-dimensional sewn dynamical system. We have constructed
the phase portrait which represents the global dynamics. From the
construction, one gets all evolutionary paths admissible for all initial conditions. The set of sewn trajectories is a critical
point located at the infinity ($\dot{a}=\infty$, $a=a_{\text{fsing}}$). The trajectories pass through this critical point. On the phase
portrait the trajectory of the flat model ($k=0$) divides trajectories in the phase space on the domains occupied by closed ($k=+1$) and
open ($k=-1$) models. From the phase portrait one can derive conclusion that while all open models possess freeze singularity there are also presented
bouncing models and closed ones with initial and final singularities without the freeze singularity. The latter class of models without freeze
singularity is also generic. Note that there exists also a class of generic models without the initial singularity.

We have also classified all solutions for the negative $\Omega_{\gamma}$. This case is favored by the astronomical data \cite{Borowiec:2015qrp}. The classification
is performed in both configurational and phase spaces. From the classification one can conclude that that there is a generic class of cosmological
models in which the Big Bang singularity is replaced by the bounce.

For the case of negative $\gamma$ the phase space is in the form of two disjoint domains without any causal communication between them. The simple method of the avoidance
the domain on the left-side of line $\{b=0\}$ is to assume that $f'(R)>0$. For our model this means that $b>0$ at the very beginning. The domain of the phase space for
a negative value of $b$ can be interpreted as a region occupied by trajectories with a negative constant coupling of matter and scalar field.

The final conclusion is that the extended cosmology with $\gamma R^2$ term in the Palatini formalism has status sewn dynamical systems from the point of view dynamical
systems theory. A point of sewing represents the freeze singularity for $\gamma > 0$ or sudden singularity in the opposite case. From the
cosmological point of view a sudden singularity is represented by a bounce while a freeze one is represented by an inflationary phase.

The type of singularities in the Starobinsky model in the Palatini formalism crucially depends on the sign of the parameter $\gamma$. If $\gamma$ is positive then we obtain new singularity apart from the Big-Bang singularity. This type singularity is similar formally to the singularity of type III but has a more complex nature, because is composed de facto with two finite scale factor singularities of type III (in a past and a future). The behavior of the scale factor near this type singularity can be obtained from the expansion of function $t=t(a)$ near an inflection point.
\begin{equation}
t-t_s \simeq \pm \left. \frac{1}{2} \frac{d^2 t}{da^2} \right|_{a=a_{\text{sing}}} (a-a_{\text{sing}})^2.
\end{equation}
Therefore,
\begin{equation}
a-a_{\text{sing}} \propto
\begin{cases}
- (t_{\text{sing}} - t)^{1/2} \quad \text{for} \quad t \to t^{-}_{\text{sing}} \\
+ (t - t_{\text{sing}})^{1/2} \quad \text{for} \quad t \to t^{+}_{\text{sing}}.
\end{cases}
\end{equation}
The acceleration $p=\ddot{a}$ goes to $+\infty$ as $t \to t^{-}_{\text{sing}}$ and $-\infty$ as $t \to t^{+}_{\text{sing}}$. In the consequence $p$ is undefinite because of the lack of continuity in this point of gluing. Note that the system under consideration is an example of a piecewise-smooth dynamical systems \cite{diBernardo:2008psd}.

Our general conclusion is that the presence of a sudden singularity in the model falsifies the the negative case of $\gamma$ in the Palatini cosmology. The agreements of the physics which imply evolutional scenario of the universe with the observations requires that the value of $\Omega_{\gamma}$ parameter should be extremely small beyond the possibilities of contemporary observational cosmology. From the statistical analysis of astronomical observations, we deduce that the case of negative values of $\Omega_\gamma$ can be rejected.

There are in principle two interpretation of obtained results. In the first interpretation, we treated singularities as artifacts of the Palatini variational principle which in consequence limits its application. However there is also another interpretation. We in consideration of MGT take the simplest, quadratic correction of general relativity. Of course we do not know the exact form of $f(R)$ and our theory plays the role of some kind effective theory. It is possible that if additional terms in a Taylor expansion be included these singularities disappear in a natural way. It seems to be interesting to build cosmology in the Palatini formalism under higher order terms (with respect to the Ricci scalar $R$ or its inverse $R^{-1}$) in a Taylor expansion for checking whether above mentioned singularities can appear during the cosmic evolution.

\acknowledgments{
The work has been supported by Polish National Science Centre (NCN), project DEC-2013/09/B/ST2/03455. AW acknowledges financial support from 1445/M/IFT/15. We are especially grateful to Orest Hrycyna for discussion and remarks.}

\providecommand{\href}[2]{#2}\begingroup\raggedright\endgroup

\end{document}